\documentclass[reprint,prl]{revtex4-1}
\usepackage{bm,epsfig,graphicx,color,amssymb,amsmath}

\begin{document}

\title{Cooperative amplification of energy transfer in plasmonic systems}


\author{Vitaliy N. Pustovit and Augustine M. Urbas}  

\affiliation{Materials and Manufacturing Directorate, Air Force Research Laboratory, WPAFB, OH 45433, USA}

\author{Tigran V. Shahbazyan}
\affiliation{Department of Physics, Jackson State University, Jackson, MS 39217, USA}


\begin{abstract}
We study cooperative effects in energy transfer (ET) from an ensemble of donors to an acceptor near a plasmonic nanostructure. We demonstrate that in cooperative regime ET takes place from plasmonic superradiant and subradiant states rather than from individual donors leading to a significant increase of ET efficiency. The cooperative amplification of ET relies on the  large coupling of superradiant states to external fields and on the slow decay rate of subradiant states. We show that superradiant and subradiant ET mechanisms are efficient in different energy domains and therefore can be utilized independently. We present numerical results demonstrating the amplification effect for a layer of donors and an acceptor on a spherical plasmonic nanoparticle.
\end{abstract}


\maketitle


Efficient energy  transfer (ET) at the nanoscale is one of the major goals in the rapidly developing field of plasmonics.  F\"{o}rster resonance energy transfer (FRET) \cite{forster-ap48,dexter-jcp53} between spacially separated donor and acceptor fluorophores, e.g., dye molecules or semiconductor quantum dots (QD), underpins diverse phenomena in biology, chemistry and physics such as photosynthesis, exciton transfer in molecular aggregates, interaction between proteins \cite{lakowicz-book,andrews-book} or, more recently, energy transfer between QDs and in QD-protein assemblies \cite{willard-prl01,klimov-prl02,clark-jpcc07}.  FRET spectroscopy is widely used, e.g.,  in studies of protein folding \cite{deniz-pnas00,lipman-science03}, live cell protein localization \cite{selvin-naturesb00,sekar-jcb03}, biosensing \cite{gonzalez-bj95,medintz-naturemat03}, and light harvesting \cite{andrews-lp11}.

During past decade, significant advances were made in ET enhancement and control by placing molecules or QDs in microcavities \cite{hopmeier-prl99,andrew-science00,finlayson-cpl01} or near plasmonic materials such as metal films and nanoparticles (NPs) \cite{leitner-cpl88,lakowicz-jf03,andrew-science04,lakowicz-jpcc07-1,lakowicz-jpcc07-2,krenn-nl08,rogach-apl08,bodreau-nl09,yang-oe09,an-oe10,lunz-nl11,zhao-jpcc12,west-jpcc12,lunz-jpcc12}. While  F\"{o}rster transfer  is efficient only for relatively short donor-acceptor separations  $\sim$10 nm \cite{lakowicz-book}, a plasmon-mediated transfer channel supported by metal NPs \cite{nitzan-cpl84,nitzan-jcp85,druger-jcp87,dung-pra02,stockman-njp08,pustovit-prb11},  films and waveguides \cite{dung-pra02,moreno-nl10} or doped monolayer graphene \cite{velizhanin-prb12}, can significant increase the transition rate at larger distances between  donor and  acceptor. At the same time, dissipation in metal and plasmon-enhanced radiation  reduce the fraction of donor's energy available for transfer to the acceptor.  In a closely related phenomenon of plasmon-enhanced fluorescence from a single fluorophore \cite{feldmann-nl05,novotny-prl06,sandoghdar-prl06,halas-nl07}, the interplay between dissipation and radiation channels, which determines fluorophore's quantum efficiency, depends sensitively on its distance to the metal surface \cite{nitzan-jcp81,ruppin-jcp82}. A  nearby acceptor  will absorb some of the donor fluorophore energy via three main transfer channels:   F\"{o}rster channel, non-radiative plasmon-mediated channel, and  plasmon-enhanced radiative channel, the latter being dominant for intermediate distances \cite{pustovit-prb11}. The fraction of the donor energy absorbed by the acceptor is then determined by an interplay between transfer, radiation and dissipation channels, so that an increase of ET efficiency implies either increase of the transfer rate or reduction of the dissipation and/or radiation rates.

\begin{figure}[tb]
 \begin{center}
  \includegraphics[width=0.85\columnwidth]{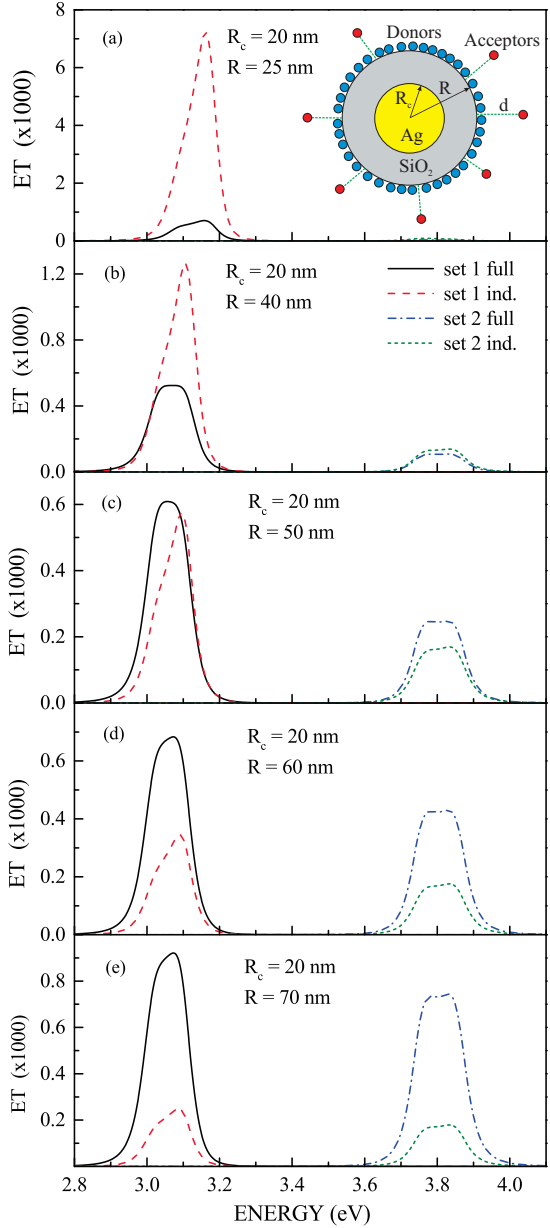}
  \end{center}
  \caption{\label{fig:rad20} ET for $N=100$ donors on top of spherical core-shell NP with Ag core radius $R_{c}=20$ nm and SiO$_{2}$ shell thickness 5 nm (a), 20 nm (b), 30 nm (c), 40 nm (d), and 50 nm (e), and acceptor at a distance $d=10$ nm from NP surface. Full calculations for two donor-acceptor sets with spectral  bands tuned to dipole (set 1) and high-\textit{l} (set 2) plasmon resonances are compared to individual donors approximation.
  }
  \end{figure}

Here we describe a novel \textit{cooperative amplification} mechanism for ET from an \textit{ensemble} of donors to acceptors that takes advantage of the subtle balance between energy flow channels in a plasmonic system. In a typical experimental setup, a large number of donors are deposited on top of a silica shell around a gold or silver core of a spherical core-shell NP, while the acceptors are attached to the NP surface via linker molecules (see schematic in Fig.~\ref{fig:rad20}). 
If the donors' separation from the metal surface is not so small that dissipation is the dominant channel, then the donors' coupling through NP plasmons gives rise to new system eigenstates -- superradiant and subradiant states \cite{dicke-pr54}, which, in the presence of a NP, are considerably more robust due to a strong plasmonic enhancement of radiative coupling \cite{pustovit-prl09,pustovit-prb10}.  In this case, ET to an acceptor takes place from these collective states rather than from each of many individual donors. Importantly, the energy flows in a system in the cooperative regime differ dramatically from those in a system of individual donors. While a few superradiant states carry only a small fraction of donors' energy, the  large matrix element with external electric fields leads to a huge decay rate that scales with the system size. In a similar manner, the large coupling of superradiant states with the electric field of an acceptor spatially separated from the donor layer increases the transfer rate and ensures, as we  demonstrate below, a much more efficient plasmon-assisted ET than from the same number of individual donors. 

%

On the other hand, the multitude of subradiant states which store the nearly entire system energy are characterized by a much slower decay rate than individual donors coupled to a NP \cite{pustovit-prl09,pustovit-prb10} implying lower energy losses through dissipation and radiation channels.   As we show below, this reduction of losses in the cooperative regime leads to a dramatic ET amplification relative to ET from individual donors. Importantly, here the reduction of losses does not require continuous pumping to sustain loss compensation by gain medium, but takes place "naturally" due to plasmon exchanges by the donors. 

To pinpoint the origin of cooperative amplification, consider ET from a single donor to an acceptor near a metal nanostructure. The fraction of transferred energy, $W_{ad}$, in unit frequency interval relative to the total energy stored in the donor, $W_{d}$, is given by \cite{pustovit-prb11}
\begin{align}
\label{fret-new}
\frac{1}{W_{d}}\frac{dW_{ad}}{d\omega}=\frac{9\tilde{\sigma}_{a}(\omega) }{8\pi k^4} \,
 \tilde{f}_{d}(\omega) \left |\tilde{D}_{ad}(\omega)\right |^{2}\dfrac{\gamma_{d}^{r}}{\Gamma_{d}(\omega)},
\end{align}
where $\gamma_{d}^{r}$ is the donor \textit{free space} radiative decay rate \cite{novotny-book}, $\Gamma_{d}$ is its \textit{full} decay rate that depends on its distance to the metal surface, $\tilde{f}_{d}$ and $\tilde{\sigma}_{a}$ are, respectively, donor's spectral function and aceptor's absorption cross section  modified  by the metal, $\tilde{D}_{da}$ is the donor-acceptor coupling that includes direct Coulomb as well as plasmon-mediated channels, and $k=\omega/c$ is light wavevector. The ET efficiency is mainly governed by the competition between the factor $|\tilde{D}_{ad}(\omega) |^{2}$ that determines plasmon-enhanced transition rate  and the quenching factor $\gamma_{d}^{r}/\Gamma_{d}$.  In the absence of metal, only Coulomb interaction contributes to the transition rate, i.e., $|D_{ad}|^{2}\propto r_{ad}^{-6}$, where $r_{ad}$ is donor-acceptor separation, and there is no quenching, i.e., $\gamma_{d}^{r}/\Gamma_{d}=1$,  so that frequency integration of Eq.~(\ref{fret-new}) yields the standard F\"{o}rster transfer rate $\left(r_{F}/r_{ad}\right)^{6}$, where $r_{F}$ is the F\"{o}rster radius determined by the overlap of donor and acceptor spectral bands. In the case of many individual donors (i.e., not interacting with each other), the r.h.s of Eq.~(\ref{fret-new})  is summed over all donors positions.


For an ensemble of donors, ET takes place from the system eigenstates, hereafter labeled  by $J$, and  Eq.~(\ref{fret-new}) holds for each eigenstate, while the ET spectral density is obtained by summation over all eigenstates $J$,
\begin{equation}
\label{dw-final}
\frac{1}{W_{d}}\dfrac{dW_{\rm ens}}{d\omega}
=\frac{9\tilde{\sigma}_{a}(\omega)}{8 \pi k^4}\sum_{J}\tilde{f}_{J}(\omega)\left |\tilde{D}_{aJ}(\omega)\right |^{2} \dfrac{\gamma_{d}^{r}}{\Gamma_{J}(\omega)},
\end{equation}
where $\Gamma_{J}$, $\tilde{f}_{J}(\omega)$, and $\tilde{D}_{aJ}(\omega)$ are, respectively, the eigenstate $J$ decay rate, spectral function, and coupling strength to the acceptor.  The full energy absorbed by an acceptor is obtained by frequency integration of Eq.~(\ref{dw-final}). The derivation of Eq.~(\ref{dw-final}) is provided in Supplemental Material, and here we discuss its implications and present our numerical results.

%
  \begin{figure}[tb]
 \begin{center}
  \includegraphics[width=0.85\columnwidth]{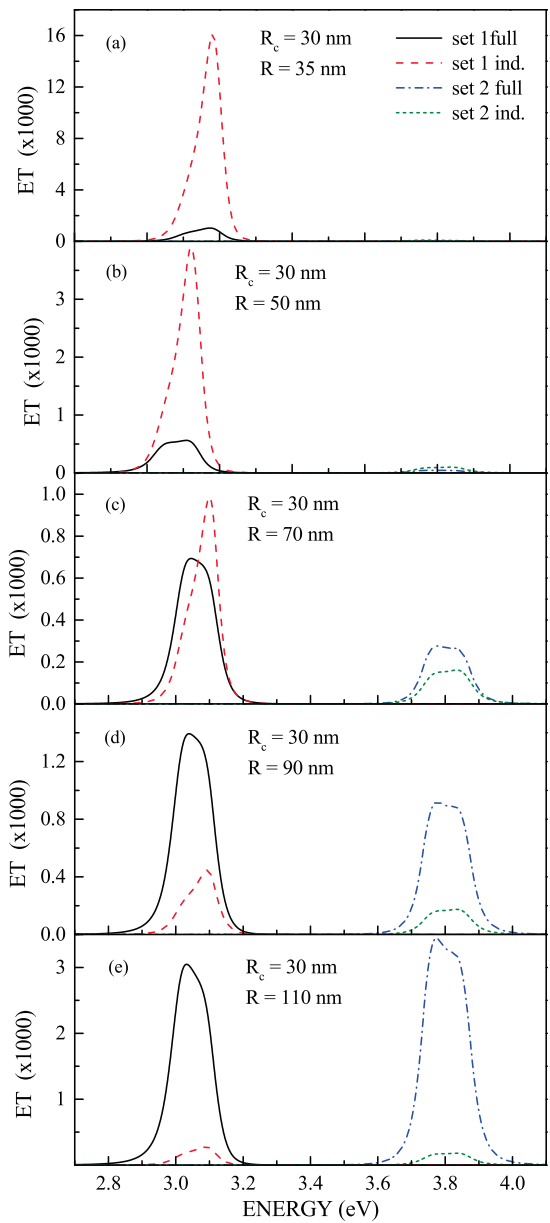}
  \end{center}
  \caption{\label{fig:rad30} Same as Fig.~\ref{fig:rad20} for core radius $R_{c}=30$ nm and shell thickness 5 nm (a) and 20 nm (b), 40 nm (c), 60 nm (d), and 80 nm (e).
  }
  \end{figure}

In the cooperative regime,  the superradiant states are strongly coupled to the electric field of an acceptor, i.e., $D_{aJ}\gg D_{ad}$, while subradiant states decay much slower than individual donors, i.e., $\Gamma_{J}\ll \Gamma_{d}$. In both cases, the result is a significant amplification of ET.  This is  illustrated in Figs.~\ref{fig:rad20} and \ref{fig:rad30} for an ensemble of 100 donors randomly distributed on surface of a spherical core-shell NP of radius $R$ immersed in water, with Ag core radius $R_{c}$  and Silica shell thickness $L=R-R_{c}$, and an acceptor at distance $d=10$ nm from the NP surface (see inset in Fig.~\ref{fig:rad20}). We assume that the donors' and acceptor's dipole orientations are all normal to the NP surface and their respective emission and absorption bands are Lorentzians of width 0.1 eV centered at energies $\omega_{d}$ and $\omega_{a}$. We expect that in the cooperative regime, the superradiant states are dominant at energies near the dipole plasmon resonance, while subradiant states are best developed at energies close to those of $l=2$ and $l=3$ plasmons (note that, at a given distance, dipole-NP interaction rapidly falls with increasing $l$). Accordingly, we use two sets of donors and acceptors:  set 1 has $\omega_{d}$  lying in the dipole plasmon band (at 3.15 eV), and set 2 has $\omega_{d}$  lying in the higher-\textit{l} plasmons region (at  3.85 eV); in both cases, $\omega_{a}$ is redsifted by 0.1 eV from the corresponding $\omega_{d}$. Note that high-\textit{l} plasmons with energies above 4.0 eV are damped by electronic interband transitions in Ag.  In all calculations, we used experimental Ag permittivity and included angular momenta up to $l_{\rm max}=75$.

%
  \begin{figure}[tb]
 \begin{center}
  \includegraphics[width=0.85\columnwidth]{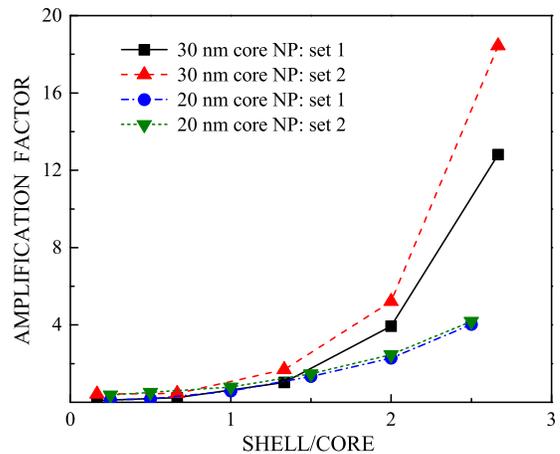}
  \end{center}
  \caption{\label{fig:ampl} Amplification factor of frequency-integrated ET relative to that for individual donors vs. shell/core ratio.
  }
  \end{figure}

In Fig.~\ref{fig:rad20} we plot the energy dependence of normalized ET, given by Eq.~(\ref{dw-final}), for Ag core radius $R_{c}=20$ nm and SiO$_{2}$ shell thickness in the range from 5 nm to 50 nm (i.e., overall NP radius $R$ between 25 and 70 nm). ET for each donor-acceptor set is compared to ET calculated for individual donors and all curves are normalized per donor. For relatively thin shells, the individual donor approximation yields significantly higher ET for set 1 since it neglects plasmon exchanges between the donors and hence underestimates the dissipation, while ET for set 2 is suppressed by very strong dissipation in the high-\textit{l} plasmon region. With increasing shell thickness, as donors move away from the metal core, the system transitions to the cooperative regime \cite{pustovit-prl09,pustovit-prb10} and ET from \textit{superradiant} states (set 1) overtakes ET from individual donors due to the much stronger coupling to the acceptor, as discussed above. At the same time, ET from \textit{subradiant} states emerges (set 2) and, for thicker shells, significantly exceeds ET from individual donors due to the much slower decay rates. Remarkably, the superradiant and subradiant  amplification mechanisms can be utilized independently  in different energy domains.

For larger NP, the ET amplification sharply increases (see Fig.~\ref{fig:rad30})  due to a stronger plasmon-enhanced radiative coupling between the donors \cite{pustovit-prb11} that leads to a more robust cooperative regime \cite{pustovit-prl09,pustovit-prb10}. The peak enhancement for both set 1 and set 2 relative to individual donors reaches $\sim 10$ for the largest NP. The crossover to the cooperative regime is determined by the ratio of shell thickness to core radius  ($L/R_{c}$) rather than by shell thickness alone, and therefore ET  for $R_{c}=30$ nm core NP overtakes ET from individual donors at thicker shells (compare Figs.~\ref{fig:rad20} and \ref{fig:rad30}).  Importantly, the evolution of  cooperative  ET  and of ET from individual donors with increasing shell thickness show opposite  trends: the former \textit{increases} with thickness while the latter is reduced. The role of cooperative effects is most pronounced in frequency-integrated ET relative to that for individual donors (amplification factor). This shows a dramatic ET increase for larger NP at similar values of shell/core ratio (Fig.~\ref{fig:ampl}). The onset of the cooperative regime corresponds to amplification factor $\sim 1$ and takes place at shell/core ratio $\sim 1$.  With increasing shell thickness, amplification factor reaches $\sim 5$ for 20 nm core NP and  $\sim 20$ for 30 nm core NP. 

With further increase of NP size, ET amplification should eventually  saturate and, as the system size exceeds the radiation wavelength $\lambda$ ($\sim 400$ nm in our case), scale back to $\sim 1$. Indeed, as system size approaches $\lambda$, the plasmonic field enhancement  weakens due to retardation effects and the cooperative regime is destroyed by the dissipation effects \cite{pustovit-prl09,pustovit-prb10}. The optimal NP size for cooperative  amplification of ET is expected to be similar to that for plasmon-enhanced fluorescence  \cite{feldmann-nl05,novotny-prl06,sandoghdar-prl06,halas-nl07}.


 \begin{acknowledgments}
This work was performed while the first author held a National Research Council Research Associateship Award at AFRL. Support by AFRL Materials and Manufacturing Directorate Applied Metamaterials Program is also acknowledged. Work at JSU was supported through NSF under Grant DMR-1206975, CREST Center, and EPSCOR program.
\end{acknowledgments}

{}
\appendix

\section*{Supplemental Material}

Here we present the derivation of  Eq. (2) by extending our single-donor model of ET \cite{pustovit-prb11} to the case $N$ donors and a single acceptor near a plasmonic nanonostructure following our approach to plasmon-assisted cooperative effects \cite{pustovit-prl09,pustovit-prb10}. The donors and the acceptor are described by point-like dipoles with excitation frequencies $\omega_{d}$ and $\omega_{a}$, respectively,  located at ${\bf r}_{j}$ ($j=1,...,N$) and  ${\bf r}_{a}$ with induced dipole moments ${\bf p}_j(\omega )=p_j(\omega) {\bf e}_{j}$ and ${\bf p}_{a}(\omega )=p_{a}(\omega) {\bf e}_{a}$  along ${\bf e}_{j}$ and ${\bf e}_{a}$. All dipoles are driven by the common electric field,
\begin{align}
\label{moments3}
&{\bf p}_{a} (\omega ) =\alpha_{a} (\omega ) {\bf E}({\bf r}_{a}, \omega), \\
&{\bf p}_{j} (\omega ) =\alpha_{j} (\omega ) {\bf E}({\bf r}_{j}, \omega)+{\bf p}_{j}^{0}(\omega),
\nonumber
\end{align}
where $\alpha_{a} (\omega )=\alpha'_{a} (\omega )+i\alpha''_{a} (\omega )$ and $\alpha_{j} (\omega )=\alpha'_{j} (\omega )+i\alpha''_{j} (\omega )$ are fluorophores complex polarizabilities, ${\bf p}_{d}^{0}(\omega ) = \alpha_{d} (\omega ) {\bf e}_{j} E_{0}e^{i\varphi_{j}}$ are the donors initial dipole moments with some constant $E_{0}$ and random phases $\varphi_{j}$ simulating incoherent excited states. The electric field $\bf E$ is a solution of Maxwell's equation with dipole sources \cite{novotny-book}
\begin{equation}
\label{field2}
{\bf E}({\bf r},\omega ) =  \frac{4\pi \omega^2}{c^2} \sum_{\beta}  {\bf G}( {\bf r}, {\bf r}_{\beta}; \omega) \cdot {\bf p}_{\beta} (\omega ),
\end{equation}
where index $\beta=(a,j)$ runs over \textit{all} dipoles positions,  ${\bf G}({\bf r},{\bf r}';\omega)$ is Maxwell's equation Green's dyadic, satisfying ${\bm \nabla} \times {\bm\nabla} \times \hat{\bf G} - \epsilon({\bf r},\omega) (\omega/c)^{2}\hat{\bf G} = \hat{\bf I}$, and $\epsilon({\bf r},\omega)$ is dielectric function of the metal, $\epsilon(\omega)$, if {\bf r} resides inside the metal region and that of the outside medium, $\epsilon_{0}$, otherwise. We are interested in the energy absorbed by the acceptor in unit frequency interval,
\begin{equation}
\label{dw3} 
\frac{dW_{ad}}{d\omega}= -\frac{\omega}{\pi}\, {\rm Im} \left [{\bf p}^{*}_a(\omega) \cdot {\bf E}({\bf r}_a,\omega)\right ]=\frac{\omega \alpha''_{a}}{\pi}\left |\frac{p_{a}}{\alpha_{a}}\right |^{2},
\end{equation}
where we used ${\bf E}({\bf r}_a,\omega)={\bf p}_{a}(\omega)/\alpha_{a}(\omega)$ from Eq.~(\ref{moments3}). A closed system for $p_{j}(\omega)$ is obtained by using Eq.~(\ref{field2}) to eliminate the electric field from Eq.~(\ref{moments3}),
\begin{align}
\label{moments4}
 &p_{a}  + \alpha_{a}  D_{aa} p_{a}  +\alpha_{d} \sum_{j}  D_{aj} p_{j}  
= 0,\\
 &p_{j}  +\alpha_{d}  D_{ja}  p_{a} + \alpha_{d} \sum_{k}  D_{jk} p_{k}  
= p_{j}^{0},
\nonumber
\end{align}
where we introduced frequency-dependent matrix 
\begin{equation}
\label{d-matrix1}
D_{\beta\beta'} (\omega ) = -\frac{4\pi \omega^2 }{c^2} 
{\bf e}_{\beta} \cdot {\bf G}({\bf r}_{\beta},{\bf r}_{\beta'};\omega) \cdot {\bf e}_{\beta'} .
\end{equation}
Solving Eq.~(\ref{moments4}), we find  
\begin{equation}
\label{pa}
p_{a}=-\dfrac{\alpha_{a}}{1+\alpha_{a}\tilde{D}_{aa}}\sum_{jk}D_{aj}S_{jk}p_{k}^{0},
\end{equation}
where the matrix $S_{jk}$ is defined through its inverse as
\begin{equation}
S_{jk}^{-1}(\omega)=\delta_{jk}+\alpha_{d}D_{jk}(\omega)
\end{equation}
and $\tilde{D}_{aa}$ describes self-interaction of the acceptor near a nanostructure in the presence of donors,
\begin{equation}
\tilde{D}_{aa}=D_{aa}-\alpha_{d}\sum_{jk}D_{aj}S_{jk}D_{ka}.
\end{equation}
Substituting $p_{a}$ from Eq.~(\ref{pa}) into Eq.~(\ref{dw3}) and averaging over random phases $\varphi_{j}$, we obtain
\begin{align}
\label{dw4}
\frac{dW_{}}{d\omega}
=\dfrac{\omega E_{0}^{2}}{\pi}\, \dfrac{\left |\alpha_{d}\right |^{2}\alpha''_a }{\left| 1+\alpha_a \tilde{D}_{aa} \right|^2}\sum_{jkl}D_{aj}S_{jk}S_{kl}^{\dagger}D_{la}^{\dagger}.
\end{align}
The system eigenstates are obtained by diagonalizing the coupling matrix of donors, $D_{jk}$. The eigenstates $|J\rangle$ satisfy equation $\hat{D}|J\rangle=D_{J}|J\rangle$, where the eigenvalues $D_{J}$ can be presented in  the form 
\begin{equation}
\label{eigenvalue}
D_{J}=\frac{2}{3}k^{3}\left (\Delta_{J}-i\Gamma_{J}\right )/\gamma_{d}^{r}.
\end{equation}
Here $\Delta_{J}$ and $\Gamma_{J}$ are, respectively, frequency shift and decay rate of the eigenstate $J$ while $\gamma_{d}^{r}=\frac{2}{3}k^{3}\mu_{d}^{2}$ is donors free space radiative decay rate ($\mu_{d}$ is dipole matrix element). We now introduce the acceptor coupling to the eigenstate $J$,
\begin{equation}
D_{aJ}=\sum_{j}D_{aj}\langle j|J\rangle,
\end{equation}
where $\langle j|J\rangle$ stands for $j$th element of eigenvector $J$, and the dressed eigenstates and acceptor polarizabilities,
\begin{equation}
\label{polar-dress1}
\tilde \alpha_{J}= \frac{\alpha_{d}}{1+ D_{J} \alpha_{d}},
~~
\tilde \alpha_{a}= \frac{\alpha_{a}}{1+ D_{aa}\alpha_{a}}.
\end{equation}
The dressed eigenstates polarizabilities satisfy the relation
\begin{equation}
\label{optical1}
\tilde{\alpha}''_{J} + D''_{J} |\tilde{\alpha}_{J}|^2 = \frac{\alpha''_{d}}{\left|1+ D_{J} \alpha_{d}\right|^2},
\end{equation}
which expresses energy balance of an eigenstate between total extinction described by $\tilde{\alpha}''_{j}$, external losses such as radiation and dissipation in metal encoded in  $D''_{jj}(\omega)$,  and absorption in the presence of environment (r.h.s.). This relation generalizes a similar relation for a single donor \cite{pustovit-prb11} to the case of an ensemble. For high-yield donors ($\alpha''_{d}=0$),  Eq. (\ref{optical1}) implies the optical theorem in absorptive environment for each system eigenstate,
\begin{equation}
\label{optical4}
\tilde{\alpha}''_{J} =-D''_{J}|\tilde{\alpha}_{J}|^{2}=\frac{2}{3}k^{3}|\tilde{\alpha}_{J}|^{2}\frac{\Gamma_{J}}{\gamma_{d}^{r}}.
\end{equation}
The radiated energy of an isolated donor, $W_{d}$, reads \cite{pustovit-prb11}
\begin{equation}
\label{Wd1}
W_{d}=\frac{E_{0}^{2}}{\pi}\int d\omega \omega  \tilde {\alpha}''_{d0} (\omega),
\end{equation}
where $\tilde {\alpha}_{d0} =\alpha_{d} \left (1-i\frac{2}{3}k^{3}\alpha_{d}\right )^{-1}$ is the donor polarizability in the radiation field.
Using the optical theorem Eq.~(\ref{optical4}) and normalizing Eq.~(\ref{dw4}) to $W_{d}$, we obtain 
\begin{equation}
\label{dw-final1}
\frac{1}{W_{d}}\dfrac{dW_{ad}}{d\omega}
=\frac{9\tilde{\sigma}_{a}(\omega)}{8 \pi k^4}\sum_{J}\tilde{f}_{J}(\omega)\left |\tilde{D}_{aJ}(\omega)\right |^{2} \dfrac{\gamma_{d}^{r}}{\Gamma_{J}(\omega)},
\end{equation}
where  
\begin{equation}
\tilde{D}_{aJ}(\omega)=\dfrac{D_{aJ}}{1-\tilde{\alpha}_{a}\sum_{J}D_{aJ}\tilde{\alpha}_{J}D_{Ja}}
\end{equation}
is the acceptor-eigenstate coupling that includes high-order transitions, and
\begin{equation}
\label{sigma-modified2}
\bar{\sigma}_{a}(\omega)=\dfrac{4\pi k}{3}\dfrac{\alpha''_{a}(\omega)}{\left | 1+\alpha_{a} D_{aa} \right|^{2}},
~~~ 
\tilde{f}_{J}(\omega)=\dfrac{\omega \tilde{\alpha}''_{J}(\omega)}{\int d\omega \omega \tilde{\alpha}''_{d0}}
\end{equation}
are the acceptor's absorption crossection and the eigenstates spectral function. Note that the latter  is not integral-normalized to unity. Finally, integrating Eq.~(\ref{dw-final1}) over frequency, we arrive at
\begin{equation}
\label{dw-final2}
\frac{W_{ad}}{W_{d}}
=\frac{9}{8 \pi }\sum_{J}\int \frac{d\omega}{k^4}\tilde{\sigma}_{a} \tilde{f}_{J} \left |\tilde{D}_{aJ}\right |^{2}\dfrac{\gamma_{d}^{r}}{\Gamma_{J}} ,
\end{equation}
%

The coupling matrix $D_{\beta\beta'}$ can be decomposed into free space and NP-induced radiative and nonradiative parts, $D_{\beta\beta'}=D_{\beta\beta'}^{0}+D_{\beta\beta'}^{r}+D_{\beta\beta'}^{nr}$. For a spherical NP,  in long-wave limit, and for normal dipole orientations, $D_{\beta\beta'}$ takes the form \cite{pustovit-prb10}
\begin{align}
\label{d-matrix-perp1}
D_{\beta\beta'}^{0}=
&
\left [ 1 + \sin^2 \left (\theta_{\beta\beta'}/2\right )\right ]/r_{\beta\beta'}^{3},
\nonumber\\
D_{\beta\beta'}^{r}=
 &
 -i\frac{2}{3}k^{3}\left [1+ 2 \alpha_1 \left  (\frac{1} {r_{\beta}^3}+ \frac{1} {r_{\beta'}^3} \right  ) 
+ \frac{4 |\alpha_1|^2}{r_{\beta}^3 r_{\beta'}^3}\right  ](\hat{\bf r}_{\beta}\cdot \hat{\bf r}_{\beta'}),
\nonumber\\
D_{\beta\beta'}^{nr} =
&
-\sum_{l}\frac{\alpha_{l} (l+1)^2  } {r_{\beta}^{l+2} r_{\beta'}^{l+2}}P_{l} (\hat{\bf r}_{\beta}\cdot \hat{\bf r}_{\beta'}),
\end{align}
where $\alpha_{l}$ is NP polarizability, $P_{l}(x)$ is the Legendre polynomial, $\hat{\bf r}_{\beta}$ is the unit vector along radial direction pointing at $\beta$th molecule, and $\theta_{\beta\beta'}$ is the angle between $\beta$th and $\beta'$th molecules positions ($\hat{\bf r}_{\beta}\cdot \hat{\bf r}_{\beta'}=\cos\theta_{\beta\beta'} $). 

\end{document}